\documentclass[letterpaper,12pt]{article}

\title{Angle deficit \& non-local gravitoelectromagnetism\\around a slowly spinning cosmic string}

\author{Jens Boos\,\footnote{~E-mail: \href{mailto:boos@ualberta.ca}{boos@ualberta.ca}} \\
    {\small Theoretical Physics Institute, University of Alberta, Edmonton, AB T6G 2E1, Canada} }

\date{Aug 26, 2020}

\usepackage{cite}
\usepackage{epsfig}
\usepackage[english]{babel}
\usepackage{amsmath}
\usepackage{amssymb}
\usepackage{amsbsy}
\usepackage{amstext}
\usepackage{amscd}
\usepackage{amsxtra}
\usepackage{amsopn}
\usepackage[onehalfspacing]{setspace}
\usepackage{tensor}
\usepackage{mathrsfs}
\usepackage{relsize}
\usepackage{amsfonts}
\usepackage[makeroom]{cancel}
\usepackage{selinput}

\usepackage{graphics}
\usepackage[export]{adjustbox}
\usepackage{subfig}

\usepackage{youngtab}

\newcommand{\dd}{\mbox{d}}

\newcommand{\lap}{\bigtriangleup}
\newcommand{\ts}[1]{{\boldsymbol{#1}}}

\usepackage{geometry}
\usepackage{parskip}

\usepackage[colorlinks=true,citecolor=green,linkcolor=red,filecolor=cyan,urlcolor=magenta]{hyperref}

\usepackage{color}
\definecolor{mygray}{rgb}{0.5,0.5,0.5}

\usepackage{listings} \definecolor{fadedred}{rgb}{0.6,0.0,0.0}
\definecolor{fadedblue}{rgb}{0.0,0.0,0.6}

\begin{document}

\maketitle

\begin{abstract}
Cosmic strings, as remnants of the symmetry breaking phase in the Early Universe, may be susceptible to non-local physics. Here we show that the presence of a Poincar\'e-invariant non-locality---parametrized by a factor $\exp(-\Box\ell^2)$---regularizes the gravitational field and thereby changes the properties of spacetime: it is now simply connected and the angle deficit around the cosmic string becomes a function of the radial distance. Similar changes occur for the non-local gravitomagnetic field of a rotating cosmic string, and we translate these mathematical facts into the language of non-local gravitoelectromagnetism and thereby provide a physical interpretation. We hope that these insights might provide a helpful perspective in the search for traces of non-local physics in our Universe.

{\hfill \smaller \textit{file: non-local-gm-v6.tex, Aug 26, 2020, jb} }

\vspace{3cm} {\smaller Essay written for the Gravity Research Foundation 2020 Awards for Essays on Gravitation and awarded an Honorable Mention.}

\end{abstract}

\vfill

\pagebreak

\section{Introduction}

Cosmic strings have long been an active field of study \cite{Kibble:1976sj,Vilenkin:1981zs,Hindmarsh:1994re}, with their influence on the CMB being as much as 10\% \cite{Ade:2013xla}; see also Ref.~\cite{Bronnikov:2019clf}. However, stemming from a high-energy phase of our Universe, they may also serve as interesting probes for non-local physics beyond the standard $\Lambda$CDM model.

As is well known, cosmic strings may form under fairly generic conditions in a symmetry-breaking phase of a gauge theory. If the gauge group is broken from $G$ down to a smaller group $H$, the formerly gauge-equivalent vacuum state may no longer correspond to a smoothly connected group manifold and the cosmic string is topologically protected from the surrounding vacuum state.

There is another notion of topology when it comes to cosmic strings: as it turns out, the gravitational field of a straight cosmic string, described in the weak field limit of four-dimensional General Relativity, has two striking properties: first, the spacetime is locally flat. And second, there is a conical singularity at the location of the string together with an angle deficit $\delta\varphi = 8\pi\mu G/c^2$, where $\mu$ is the linear energy density along the string.

In this Essay we would like to argue the following: suppose, for a moment, that physics in the Early Universe is susceptible to non-local modifications. In particular, let us assume that non-local effects are mediated by Poincar\'e-invariant terms of the form $\exp(-\Box\ell^2)$, as they appear in various contexts in string theory \cite{Frampton:1988kr,Tseytlin:1995uq,Biswas:2005qr,Biswas:2010yx} or non-commutative geometry \cite{Spallucci:2006zj}, where the constant $\ell>0$ is called the \emph{scale of non-locality}. Theories with these form factors have been known for quite some time \cite{Efimov:1967pjn,Tomboulis:1997gg}, feature a number of appealing UV properties both in field theory and gravity, and are under active investigation \cite{Modesto:2011kw,Biswas:2011ar,Edholm:2016hbt,Boos:2018bxf,Boos:2018bhd,Buoninfante:2018mre}.

Then, we would like to ask: what is the gravitational field of a cosmic string described by a non-local infinite-derivative theory? And what are its properties, similarities, and differences to the cosmic string obtained in linearized General Relativity?

\section{A cosmic string in non-local gravity}
Choosing Euclidean coordinates $\{t,x,y,z\}$, let us describe four-dimensional Minkowski spacetime and a perturbation around it as ($c\equiv 1$)
\begin{align}
\dd s^2 &= \eta{}_{\mu\nu}\dd x{}^\mu\dd x{}^\nu = -\dd t^2 + \dd x^2 + \dd y^2 + \dd z^2 \, , \\
g{}_{\mu\nu} &= \eta{}_{\mu\nu} + h{}_{\mu\nu} \, .
\end{align}
In our approximation a cosmic string constitutes an infinitely thin energy distribution along one spatial direction, and for simplicity we shall assume our string to be straight and aligned with the $z$-axis. Its energy-momentum tensor is
\begin{align}
\label{eq:tmunu-1}
T{}_{\mu\nu} = \mu\left( \delta{}^t_\mu \delta{}^t_\nu - \delta{}^z_\mu \delta{}^z_\nu \right)\delta(x)\delta(y) \, ,
\end{align}
where $\mu>0$ is the linear energy density (or tension) of the string such that $\mu G$ is dimensionless in units where $c=1$, and observational constraints place $\mu G \lesssim 10^{-11}$ \cite{Bronnikov:2019clf}. The linearized field equations of infinite-derivative gravity take the form
\begin{align}
\label{eq:eom}
e^{-\Box\ell^2} \Box \left( h_{\mu\nu} - \frac12 h \eta{}_{\mu\nu} \right) = -16\pi G T{}_{\mu\nu} \, ,
\end{align}
where $\ell>0$ denotes the \emph{scale of non-locality}, $\Box = -\partial_t^2+\lap$ is the d'Alembert operator, and $h = \eta{}^{\mu\nu}h{}_{\mu\nu}$. In the limit of $\ell\rightarrow 0$ one readily reproduces linearized General Relativity. The energy momentum \eqref{eq:tmunu-1} does not depend on the time $t$ nor the $z$-coordinate, and is rotationally symmetric around the $z$-axis. For that reason we make the following ansatz for $h{}_{\mu\nu}$:
\begin{align}
\label{eq:ansatz}
\ts{h} &= h{}_{\mu\nu} \dd x{}^\mu \dd x{}^\nu = \psi\left( \dd t^2 + \dd z^2 \right) + \phi\left(\dd x^2 + \dd y^2\right) \, ,
\end{align}
where $\psi=\psi(\rho)$ and $\phi=\phi(\rho)$ with $\rho^2=x^2+y^2$. The field equations \eqref{eq:eom} take the form
\begin{align}
e^{-\lap\ell^2} \lap \psi = 0 \, , \quad e^{-\lap\ell^2} \lap \phi = -16\pi G\mu \delta(x)\delta(y) \, .
\end{align}
These are two-dimensional static equations. The Laplace operator has zero mode solutions, $\lap f = 0$, called harmonic functions, which here take the form $f = a + b\rho$. The constant $a$ can be absorbed into a rescaling of the original Euclidean coordinates, and we can exclude the $b{}$-term by demanding that the fields grow at most logarithmically at large distances $\rho$, but not faster. We can then introduce the two-dimensional purely spatial Green function
\begin{align}
e{}^{-\lap\ell^2}\lap \mathcal{G}_2(\rho) = -\delta(x)\delta(y) = -\frac{1}{2\pi\rho}\delta(\rho) \, ,
\end{align}
which is known analytically \cite{Boos:2018bxf,Olver:2010}:
\begin{align}
&\mathcal{G}_2(\rho) = \frac{1}{4\pi} \left[ \text{Ei}\left( -\tfrac{\rho^2}{4\ell^2} \right) - 2\ln\left(\tfrac{\rho}{\rho_0}\right) \right] \, , \quad \text{Ei}(-x) = -E_1(x) = -\int\limits_x^\infty \dd z \frac{e^{-z}}{z} \, , \quad x > 0 \, , \\
&\mathcal{G}_2(\rho\rightarrow 0) \approx \frac{1}{4\pi}\left[\gamma+\ln\left(\tfrac{\rho_0^2}{4\ell^2}\right)\right] - \frac{\rho^2}{16\pi\ell^2} + \mathcal{O}(\rho^4) \, .
\end{align}
Here, $\rho_0$ is an integration constant which is of the order of the diameter of the cosmic string \cite{Vilenkin:1981zs}, and $\gamma=0.577\dots$ is the Euler--Mascheroni constant. The gravitational field of a  cosmic string in non-local infinite-derivative gravity is then
\begin{align}
\psi = 0 \, , \quad \phi = 16\pi G\mu \mathcal{G}_2(\rho) = 4G\mu \left[ \text{Ei}\left( -\tfrac{\rho^2}{4\ell^2} \right) - 2 \ln \left(\tfrac{\rho}{\rho_0}\right) \right] \, ,
\end{align}
which is manifestly regular at $\rho\rightarrow 0$. In the limit of large distances or vanishing non-locality, $\rho/\ell\rightarrow \infty$, one recovers linearized  General Relativity as well as the corresponding gravitational field of a cosmic string,
\begin{align}
G_2(\rho) = \lim\limits_{\ell\rightarrow 0} \mathcal{G}_2(\rho) = -\frac{1}{2\pi} \ln \left( \tfrac{\rho}{\rho_0} \right) \, , \quad \phi = -8G\mu\ln \left(\tfrac{\rho}{\rho_0}\right)\, .
\end{align}

\subsection{No distributional curvature in the presence of non-locality}
At the linear level, the ansatz \eqref{eq:ansatz} has only one non-vanishing component of the Riemann tensor (up to algebraic symmetries),
\begin{align}
R_{xyxy} = \frac12 \lap \phi = -16\pi G \mu e^{\lap\ell^2} \delta(x)\delta(y) .
\end{align}
In the second equality we employed the field equations \eqref{eq:eom}. In the limit $\ell\rightarrow 0$ one reproduces the well-known distributional curvature of the cosmic string along the $z$-axis, which in turn cuts out the z-axis from the manifold and thereby affects the spatial topology. It is no longer simply connected, since equatorial loops can no longer be contracted to a single point.

For non-vanishing non-locality $\ell>0$, however, this is not the case! In fact, one may calculate
\begin{align}
\label{eq:heat-kernel}
\frac12 \lap \phi = -16\pi G \mu K_2(\rho|\ell) \, , \quad  K_2(\rho|\ell) = \frac{1}{4\pi\ell^2} e^{-\tfrac{\rho^2}{4\ell^2}} \, ,
\end{align}
where $K_2(\rho|\ell)$ is the two-dimensional heat kernel that satisfies $K_2(\rho|\ell\rightarrow 0) = \delta(\rho)/2\pi\rho$. This shows that non-locality not only smears out $\delta$-like sources, but also removes the distributional character of the curvature.

\subsection{Modification of the angle deficit}
\begin{figure}[!htb]
\centering
\subfloat[Linearized General Relativity.]
{
    \includegraphics[width=0.49\textwidth]{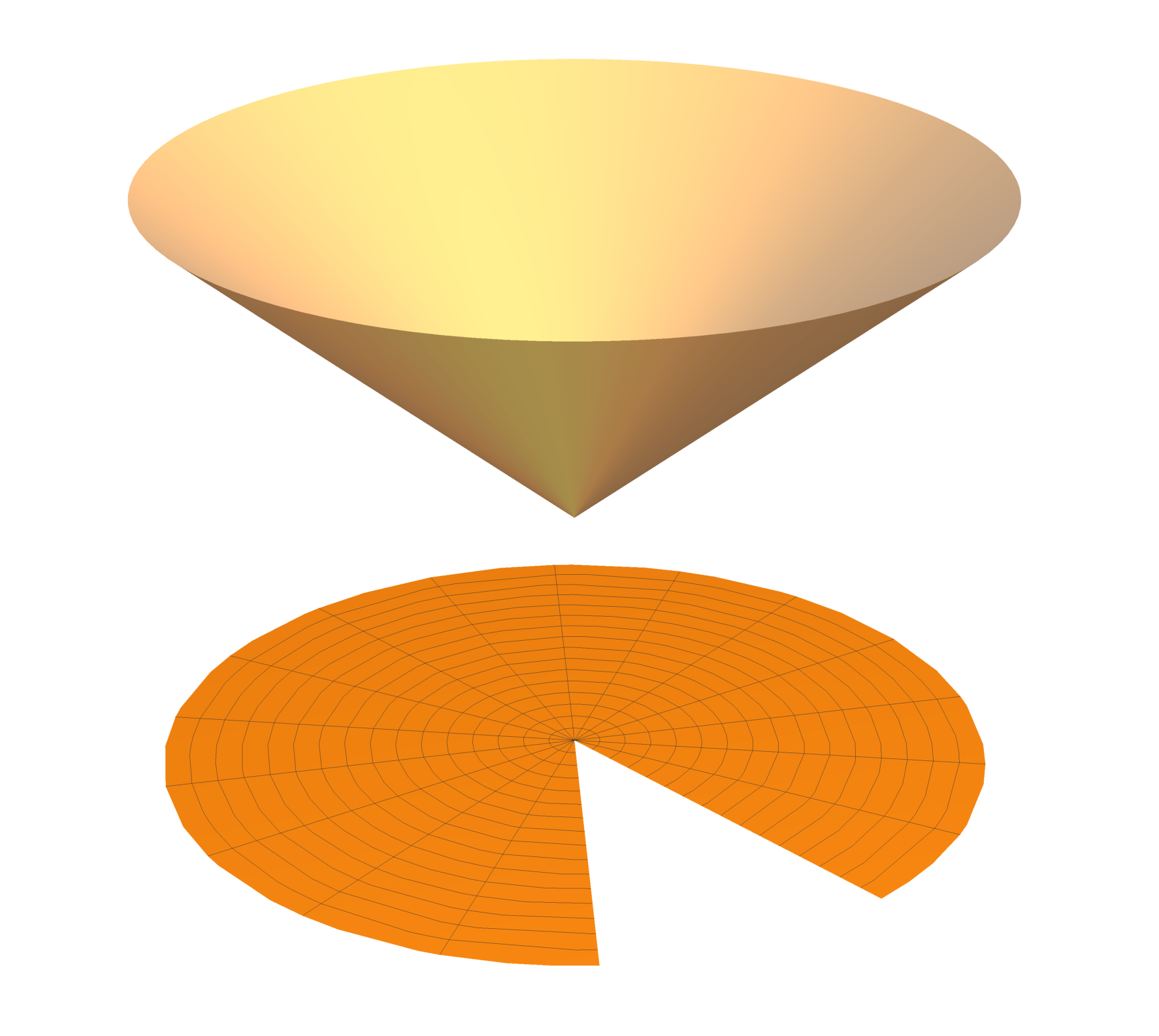}
}
\subfloat[Linearized non-local gravity.]
{
    \includegraphics[width=0.49\textwidth]{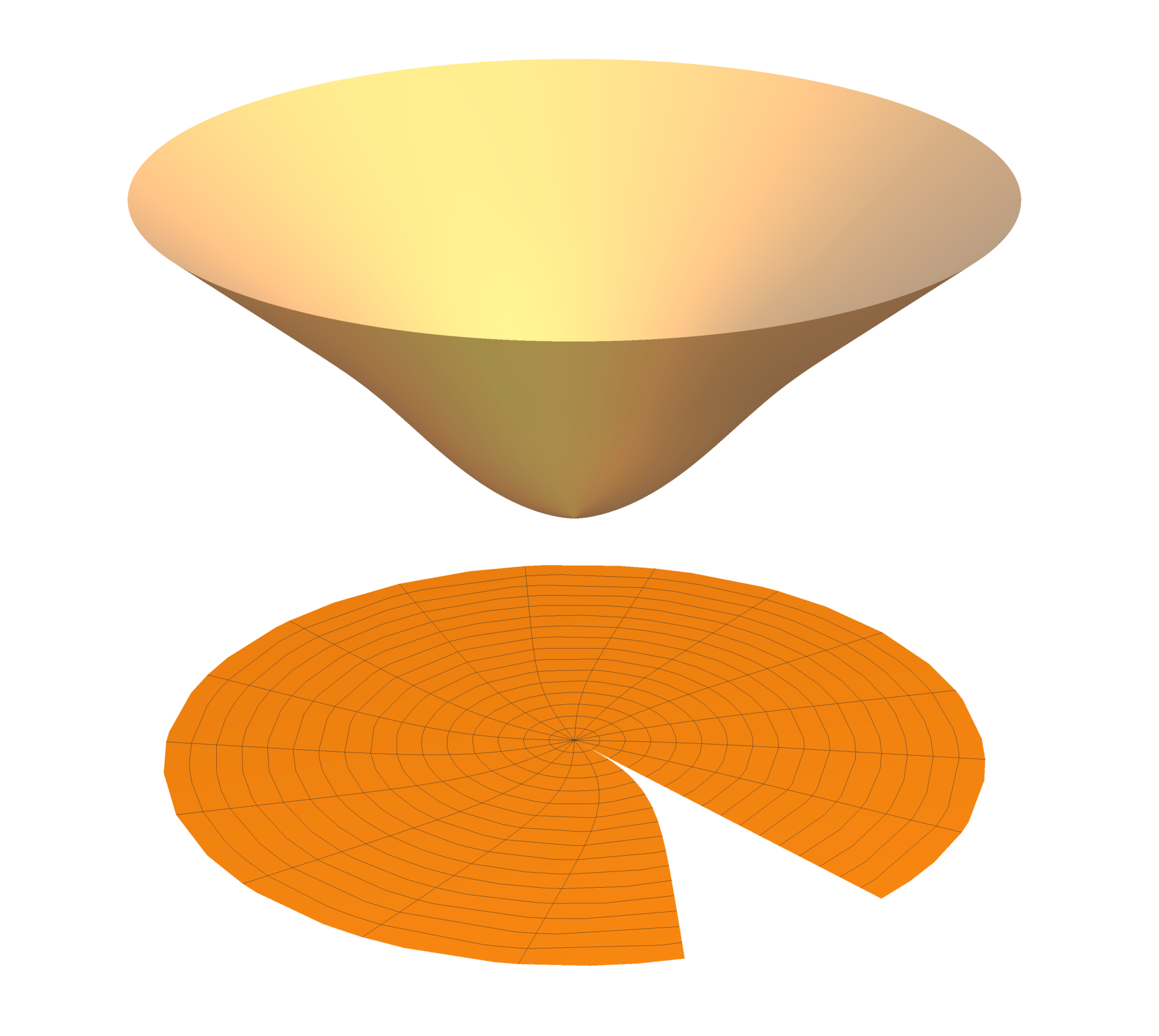}
}
\caption{The transverse spatial geometry of a cosmic string in linearized General Relativity (left) and linearized non-local gravity (right), isometrically embedded in Euclidean space. The angle deficit is constant on the left, but on the right it slowly grows from zero to the same value as in General Relativity. The resulting geometries are a cone and a smoothed cone. We used the parameters $8\mu G=0.7, \ell=0.3\rho_0$, and the circle's coordinate radius is chosen to be $2\rho_0$.}
\label{fig:cones}
\end{figure}

Let us rewrite the metric of a cosmic string in polar coordinates,
\begin{align}
\ts{g} = \left(\eta{}_{\mu\nu}+h{}_{\mu\nu}\right)\dd x{}^\mu \dd x{}^\nu = -\dd t^2 + \dd z^2 + (1+\phi)\left(\dd\rho^2 + \rho^2\dd\varphi^2\right) \, .
\end{align}
Calculating both the circumference $C(\rho)$ of a $\rho=\text{const.}$ circle and its proper radius $R(\rho)$ one finds to leading order in $\mu G$ (but for any value of $\ell$) the angle deficit
\begin{align}
\delta\varphi(\rho) = 2\pi - \frac{C(\rho)}{R(\rho)} = 8\pi G\mu\left[ 1 - \frac{\sqrt{\pi}\ell\text{erf}\left(\tfrac{\rho}{2\ell}\right)}{\rho} \right] \, .
\end{align}
In the limit of vanishing non-locality or at large distances, $\rho/\ell\rightarrow \infty$, one recovers the well-known result from General Relativity, $\delta\varphi=8\pi G \mu$. At small distances, however, the angle deficit is different, and even vanishes entirely at the location of the string:
\begin{align}
\lim\limits_{\rho\rightarrow 0} \delta\varphi(\rho) = 0 \, .
\end{align}
For this reason the local geometry in the vicinity of the cosmic string is quite regular, whereas its global properties remain unaffected: one may think of it as a cone whose tip has been smoothed out with a resulting mean curvature of $\mathcal{R} \sim -1/\ell^2$. We visualize this by employing an isometric embedding after the coordinate transformation $\rho'^2 = (1+\phi)\rho^2$; see Fig.~\ref{fig:cones} for a diagram. For more graphical representations of cosmic strings we refer to Puntigam \& Soleng \cite{Puntigam:1996vy}.

\pagebreak
\section{What about rotation?}
By linearity one may also consider a rotating cosmic string with the energy momentum
\begin{align}
T{}_{\mu\nu} = \mu\left( \delta{}^t_\mu \delta{}^t_\nu - \delta{}^z_\mu \delta{}^z_\nu \right)\delta(x)\delta(y) + \delta{}^t_{(\mu} \delta{}^i{}_{\nu)} j{}_i{}^k \partial{}_k \delta(x)\delta(y) \, ,
\end{align}
where $j_{ik} = -j_{ki}$ is the antisymmetric angular momentum tensor, and the Latin indices $i,k,\dots$ are two-dimensional indices of the $xy$-plane. The gravitational field is only modified by a gravitomagnetic potential term $\ts{A} = A_i \dd x^i$ such that the metric becomes stationary,
\begin{align}
\ts{h} = \phi\left(\dd x^2 + \dd y^2\right) + \ts{A} \dd t \, , \qquad e^{-\lap\ell^2} \lap A_i = -8\pi G j{}_i{}^k \partial_k \delta(x)\delta(y) \, .
\end{align}
The rotational symmetry in the $xy$-plane implies $\ts{A} = \ts{A}(\rho)$. Setting $j_{xy} = j$ one can show
\begin{align}
\ts{A}(\rho) = -8\pi G j\rho\,\mathcal{G}_2'(\rho)\dd\varphi = 4Gj\left( 1 - e^{-\tfrac{\rho^2}{4\ell^2}} \right) \dd \varphi \, .
\end{align}
In the limiting case of $\ell\rightarrow 0$ one reproduces the general relativistic result $\ts{A} = 4Gj\dd\varphi$, which is locally exact. Because $\varphi$ is not a continuous variable (it jumps from $2\pi$ to $0$ at one point) the field is not globally exact, however, and for that reason the magnetic field $\ts{B} = \dd\ts{A}$ is non-zero only on the $z$-axis. In general, for $\ell > 0$, the magnetic field is globally non-vanishing:
\begin{align}
\ts{B} = \dd\ts{A} = \frac{2Gj}{\ell^2} e^{-\tfrac{\rho^2}{4\ell^2}} \dd\rho \wedge \rho\dd\varphi \, .
\end{align}
Comparing with \eqref{eq:heat-kernel} one realizes that the magnetic field is proportional to the heat kernel, and for that reason becomes localized on the $z$-axis in the limiting case $\ell\rightarrow 0$. Similarly, the gravitomagnetic charge (where $\mathcal{D}$ is a disc in the $xy$-plane with radius $\rho$),
\begin{align}
Q = \int\limits_{\mathcal{D}} \ts{B} = \int\limits_{\partial\mathcal{D}} \ts{A} = 8\pi Gj\left( 1 - e^{-\tfrac{\rho^2}{4\ell^2}} \right) \, ,
\end{align}
depends on the radius $\rho$ in the non-local theory. In the limit of $\ell\rightarrow 0$, however, it becomes the constant expression $Q = 8\pi G j$. Next to the unchanged $R_{xyxy}$-component, spacetime curvature also picks up a term
\begin{align}
R_{tijk} = \partial_i \partial_{[j} A{}_{k]} = 2\partial_i B{}_{jk} \, .
\end{align}
Hence the gravitomagnetic field mimics the behavior of the gravitoelectric potential $\phi$: it is distributional in the case of General Relativity, giving rise to topological restrictions, but these restrictions disappear in the presence of non-locality.

\section{Non-local gravitoelectromagnetism}
Using the spatial Laplace--Beltrami identity $\lap = \dd \delta + \delta\dd$, where $\delta = \star\dd\star$ is the coderivative, we may re-write the field equations in a gravitoelectromagnetic (``GEM'') fashion (note that $\delta\ts{A}=0$ is just the Lorenz gauge condition, which our gravitomagnetic potential satisfies):
\begin{align}
e^{-\lap\ell^2} \dd\star \ts{E} = \ts{\rho} \, , \quad
e^{-\lap\ell^2} \dd\star \ts{B} = \ts{j} \, ,
\end{align}
which in vector calculus take the perhaps more familiar form
\begin{align}
e^{-\lap\ell^2} \ts{\nabla} \cdot \ts{E} = \rho \, , \quad
e^{-\lap\ell^2} \ts{\nabla} \times \ts{B} = \ts{j} \, .
\end{align}
In the above we defined the gravitoelectric and gravitomagnetic fields $\ts{E} = \dd \varphi$ and $\ts{B} = \dd\ts{A}$, the density $\rho = -16\pi G T_{tt}$ (and its 3-form dual $\ts{\rho} = \star \rho$), and the matter current $\ts{j} = -16\pi G \star T_{ti}\dd x{}^i$.

The structure of the GEM equations highlights the interplay of topology and non-locality. Even if the right-hand side takes the form of a $\delta$-like distribution, one may invert the non-local form factor $\exp(-\!\lap\!\ell^2)$ and bring it to the right-hand side as well. This smears any sharply concentrated matter density, implying that the usual identification of $\delta$-like sources with topological properties is no longer valid. It would be interesting to compare these findings with other recent developments in non-local gravitomagnetism by Hehl \& Mashhoon \cite{Hehl:2009es,Mashhoon:2019jkq}.

\section{Discussion and conclusions}

The Early Universe becomes more accessible with modern precision cosmological experiments under way. One possible modification may arise in form of non-local physics, which we model here by a form factor $\exp(-\Box\ell^2)$. Because these are non-zero, these theories have the advantage of not introducing any new propagating degrees of freedom at tree level \cite{Buoninfante:2018mre}, and are therefore sometimes referred to as ``ghost-free'' (as opposed to higher-derivative modifications which typically bring about new, spurious degrees of freedom that are in tension with observational cosmology).

In this Essay we studied the gravitational field of a rotating cosmic string and found notable differences from General Relativity: (i) the gravitational field is everywhere regular, (ii) the angle deficit increases with radial distance, and (iii) the gravitomagnetic field is only asymptotically exact. While these facts constitute new findings in the realm of gravitational infinite-derivative theories it remains an open question whether these non-local deviations from General Relativity necessarily lead to new observational consequences or constraints.

While an in-depth study of that question lies beyond the scope of this Essay, let us note that the non-local field equations at the heart of our considerations can be written in two equivalent ways:
\begin{align}
\label{eq:smeared-sources}
e^{-\Box\ell^2} \Box \left( h_{\mu\nu} - \frac12 h \eta{}_{\mu\nu} \right) = -16\pi G T{}_{\mu\nu} \quad
\Leftrightarrow \quad \Box \left( h_{\mu\nu} - \frac12 h \eta{}_{\mu\nu} \right) = -16\pi G T^\text{eff}_{\mu\nu} \, ,
\end{align}
where we defined an effective energy momentum tensor $T^\text{eff}_{\mu\nu}$, which in the stationary case admits an integral representation in terms of the two-dimensional heat kernel $K_2(\rho|\ell)$ defined previously:
\begin{align}
T^\text{eff}_{\mu\nu}(x) = e^{\Box\ell^2}T{}_{\mu\nu}(x) = e^{\lap\ell^2}T{}_{\mu\nu}(x) = \int\dd^2 y K_2(\sqrt{x-y}|\ell) T{}_{\mu\nu}(y) \, .
\end{align}
For this reason a linear non-local theory with a $\delta$-shaped source is equivalent to a linear local theory with a smeared matter source \cite{Giacchini:2018wlf,Briscese:2019twl}. The cosmic string geometries described in this Essay hence correspond to cosmic string solutions of linearized General Relativity supported by extended matter sources, whose width can be associated with the scale of non-locality $\ell$. Their tension $\mu G/c^2$ is therefore subject to identical constraints as in General Relativity. The \emph{difference} between General Relativity and non-local infinite-derivative gravity can make itself known in (i) time-dependent situations, where non-locality may violate causality at small scales, and in (ii) the full non-linear theory, where an equivalence of the form \eqref{eq:smeared-sources} is no longer valid.

For these reasons a full study needs to go beyond the linear level, but the results presented here make us hopeful that non-local physics, as mediated by infinite-derivative form factors, may present a promising avenue towards a well-behaved UV completion of General Relativity.

\textit{Note added in proof.}---After the submission of this Essay a paper by Kolar \& Mazumdar \cite{Kolar:2020bpo} appeared, studying the gravitational field of a spinning half-string ($z>0$) at the linear level. The solution presented in this essay has the same asymptotic form in the limit $z\rightarrow\infty$.

\section*{Acknowledgements}
I would like to thank the anonymous referee for their remarks that helped to emphasize the interpretation of non-locality in the present context, and I am grateful to Arkady A.\ Tseytlin (London) for pointing out Ref.~\cite{Tseytlin:1995uq}. I benefited from discussions with Valeri P.\ Frolov and Andrei Zelnikov (Edmonton) as well as Friedrich W.\ Hehl (Cologne), and am moreover grateful for a Vanier Canada Graduate Scholarship administered by the Natural Sciences and Engineering Research Council of Canada as well as for the Golden Bell Jar Graduate Scholarship in Physics by the University of Alberta.

\begin{singlespace}

\end{singlespace}

\end{document}